# An Environmentally Stable and Lead-Free Chalcogenide Perovskite

Tushar Gupta[1], Debjit Ghoshal[2], Anthony Yoshimura[3], Swastik Basu[1], Philippe K. Chow[4], Aniruddha S. Lakhnot[1], Juhi Pandey[5], Jeffrey M. Warrender[4], Harry Efstathiadis[6], Ajay Soni[5], Eric Osei-Agyemang[7], Ganesh Balasubramanian[7], Shengbai Zhang[3], Su-Fei Shi[2], Toh-Ming Lu[3], Vincent Meunier[3,8], and Nikhil Koratkar[1,8]*

[1]Department of Mechanical, Aerospace, and Nuclear Engineering, Rensselaer Polytechnic Institute, Troy, NY 12180, USA

[2]Department of Chemical and Biological Engineering, Rensselaer Polytechnic Institute, Troy, NY 12180, USA

[3]Department of Physics, Applied Physics, and Astronomy, Rensselaer Polytechnic Institute, Troy, NY 12180, USA

[4]U.S. Army Combat Capabilities Development Command - Armament Center - Benet Laboratories, Watervliet NY 12189, USA

[5]School of Basic Sciences, Indian Institute of Technology Mandi, Mandi 175005, HP, India

[6]Colleges of Nanoscale Science and Engineering, State University of New York Polytechnic Institute, Albany, NY 12203, USA

[7]Department of Mechanical Engineering & Mechanics, Lehigh University, Bethlehem, PA 18015, USA

[8]Department of Materials Science and Engineering, Rensselaer Polytechnic Institute, Troy, NY 12180, USA

*Corresponding author: Nikhil Koratkar

Email: koratn@rpi.edu

**Abstract**

Organic-inorganic halide perovskites are intrinsically unstable when exposed to moisture and/or light. Additionally, the presence of lead in many perovskites raises toxicity concerns. Herein is reported a thin film of $BaZrS_3$, a lead-free chalcogenide perovskite. Photoluminescence and X-ray diffraction measurements show that $BaZrS_3$ is far more stable than methylammonium lead iodide ($MAPbI_3$) in moist environments. Moisture- and light-induced degradations in $BaZrS_3$ and $MAPbI_3$ are compared by using simulations and calculations based on density functional theory. The simulations reveal drastically slower degradation in $BaZrS_3$ due to two factors – weak interaction with water, and very low rates of ion migration. $BaZrS_3$ photo-detecting devices with photo-responsivity of ~46.5 mA $W^{-1}$ are also reported. The devices retain ~60% of their initial photo-response after 4 weeks in ambient conditions. Similar $MAPbI_3$ devices degrade rapidly and show ~95% decrease in photo-responsivity in just 4 days. The findings establish the superior stability of $BaZrS_3$ and strengthen the case for its use in optoelectronics. New possibilities for thermoelectric energy conversion using these materials are also demonstrated.

**Introduction**

Over the last decade, organic-inorganic halide perovskite (OIHP) materials have taken center stage in the lively arena of optoelectronics research.[1-4] These perovskite compounds have the general chemical formula $ABX_3$ in their 3D form, where A is a small organic cation such as methylammonium ($CH_3NH_3^+$) or formamidinium ($HC(NH_2)_2^+$), B is a metal cation such as lead ($Pb^{2+}$) or tin ($Sn^{2+}$), and X is a halide ion such as iodide ($I^-$) or bromide ($Br^-$). In 2009, Kojima *et al.* introduced methylammonium lead halides ($CH_3NH_3PbX_3$) for photovoltaics.[5] Since then, OIHP materials have been the subject of many studies that have led to a broad array of applications such as solar cells,[1] light-emitting diodes,[4] and photon detectors.[2,6,7]

Despite their outstanding performance in optoelectronic applications, there are two major challenges that are inherent to these OIHP materials. The first challenge is their poor stability under common environmental conditions. Moisture is notorious for wreaking havoc on OIHP films. Methylammonium lead iodide ($CH_3NH_3PbI_3$ or $MAPbI_3$) is rapidly attacked by molecular $H_2O$, which causes it to decompose into $PbI_2$, $CH_3NH_2$ and HI.[8] Instability of OIHP materials under illumination is another significant limitation. Kim *et al.* have shown that illumination (i.e., light irradiation with energy higher than the bandgap) of $MAPbI_3$ drives iodine out of the crystal while creating iodine vacancies in the lattice.[9] This exodus of iodine, if unimpeded, can lead to the breakdown of $MAPbI_3$. Such ion migration in OIHP materials is considered an "intrinsic problem" that cannot be overcome simply by device encapsulation.[10] The second major challenge is that several members of the OIHP family contain lead (Pb), which is highly toxic and harms the environment. Specifically, $PbI_2$, the decomposition product of $MAPbI_3$, is carcinogenic. Efforts to replace Pb with Tin (Sn) in OIHP materials were initially promising but have had limited success in terms of stability.[11,12] This is primarily due to the instability of Sn in its 2+ oxidation state. In the face of these challenges, it is necessary to identify and develop lead-free perovskites that are intrinsically stable under light irradiation and when exposed to the environment.

First principles calculations have indicated that chalcogenide perovskites are promising candidates for optoelectronics.[13] These perovskites were first synthesized decades ago, but most of the previously published works focused on studying the crystal structure of these materials.[14-18] Chalcogenide perovskites are based on elements that are more environmentally friendly than Pb, but no optical nor electronic data were reported for these materials until relatively recently. Perera *et al.* reported that the bandgap of bulk barium zirconium oxysulfides could be tuned in the range of 1.75 – 2.87 eV by varying the sulfur content in $BaZr(O_xS_{1-x})_3$.[19] Niu *et al.* demonstrated iodine-catalyzed solid-state reaction as a new synthesis route and characterized the optical behavior of bulk $BaZrS_3$ and two $SrZrS_3$ polymorphs.[20] They found that these materials exhibit photoluminescence quantum efficiency and quasi-Fermi level splitting that compare well with existing high-efficiency photovoltaic materials.

Although chalcogenide perovskites have existed for decades and the basic optical properties of some of these materials are known, very little is known about chalcogenide perovskite thin films or optoelectronic devices. A detailed study of their environmental stability is also missing. Herein we report a thin-film of $BaZrS_3$ – a lead-free chalcogenide perovskite – along with an optoelectronic device based on it. We also report a quantitative comparison of $BaZrS_3$ and $MAPbI_3$ in terms of environmental stability. When compared with $MAPbI_3$, the thin film of $BaZrS_3$ showed superior stability by retaining its optoelectronic and structural character after over a month of exposure to ambient conditions. Even an aggressive environment of steam that rapidly destroyed $MAPbI_3$ was tolerated well by $BaZrS_3$. Using calculations based on density functional theory (DFT), we show that weak interactions with water and low ion-migration rates are responsible for the superior resistance of $BaZrS_3$ to water- and light-induced degradation, respectively. Our calculations provide a clear and unambiguous explanation as to why traditional perovskites ($MAPbI_3$) are unstable and why chalcogenide perovskites ($BaZrS_3$) are not. Finally, we fabricated and tested photo-detector devices using the $BaZrS_3$ thin-film material. The devices exhibited an initial mean photo-responsivity of ~46.5 mA W$^{-1}$, which decreased by ~40% after the devices were subjected to ambient conditions for 4 weeks. This is in stark contrast to similar MAPbI$_3$ photo-sensing devices, which rapidly succumbed to the environment and exhibited ~95% loss in photo-responsivity in just 4 days. Our study highlights the opportunities for further exploration of chalcogenide perovskite materials in optoelectronics. Additionally, our simulations indicate an enhancement in Seebeck coefficient and decrease in thermal conductivity with increasing temperatures for BaZrS$_3$, which could also open up exciting new applications for such materials in thermoelectric devices.

**Results and Discussion**

We used a two-step approach to deposit the chalcogenide perovskite film. In the first step, a thin film of barium zirconium oxide (BaZrO$_3$) was made by chemical solution deposition. Briefly, a solution of barium acetate, zirconium (IV) acetylacetonate and polyvinyl butyral in propionic acid was spin-coated on a 1cm × 1cm quartz substrate followed by annealing (Materials and Methods) in air to obtain the BaZrO$_3$ film. The oxide thin film appeared colorless and transparent to the eye under white light. In the second step, the oxide film was "sulfurized" by heating it under a flowing mixture of carbon disulfide and nitrogen in a tube furnace. A schematic illustration of the sulfurization process is shown in **Figure 1a**. Additional details of the process are provided in Materials and Methods. A brown film was obtained after the sulfurization step – a photograph of an oxide film alongside a sulfurized film is shown in **Figure 1b**.

An X-ray diffraction (XRD) measurement of the sulfurized film indicated that it was polycrystalline. The XRD pattern is shown in **Figure 1c**. Five significant peaks can be seen, and they align closely with the five strongest lines in the reference pattern for BaZrS$_3$ (ICDD 00-015-0327). No significant peaks for other phases were found. This indicates that the sulfurized film was BaZrS$_3$ with the orthorhombic distorted perovskite crystal structure and *Pnma* space group. The slight differences between peak and reference line positions can be attributed to residual strain due to the dissimilar substrate. The background signal in Figure 1c originates from the amorphous quartz substrate. The XRD pattern of the precursor oxide (i.e., BaZrO$_3$) film is provided in **Figure S1** (Supporting Information).

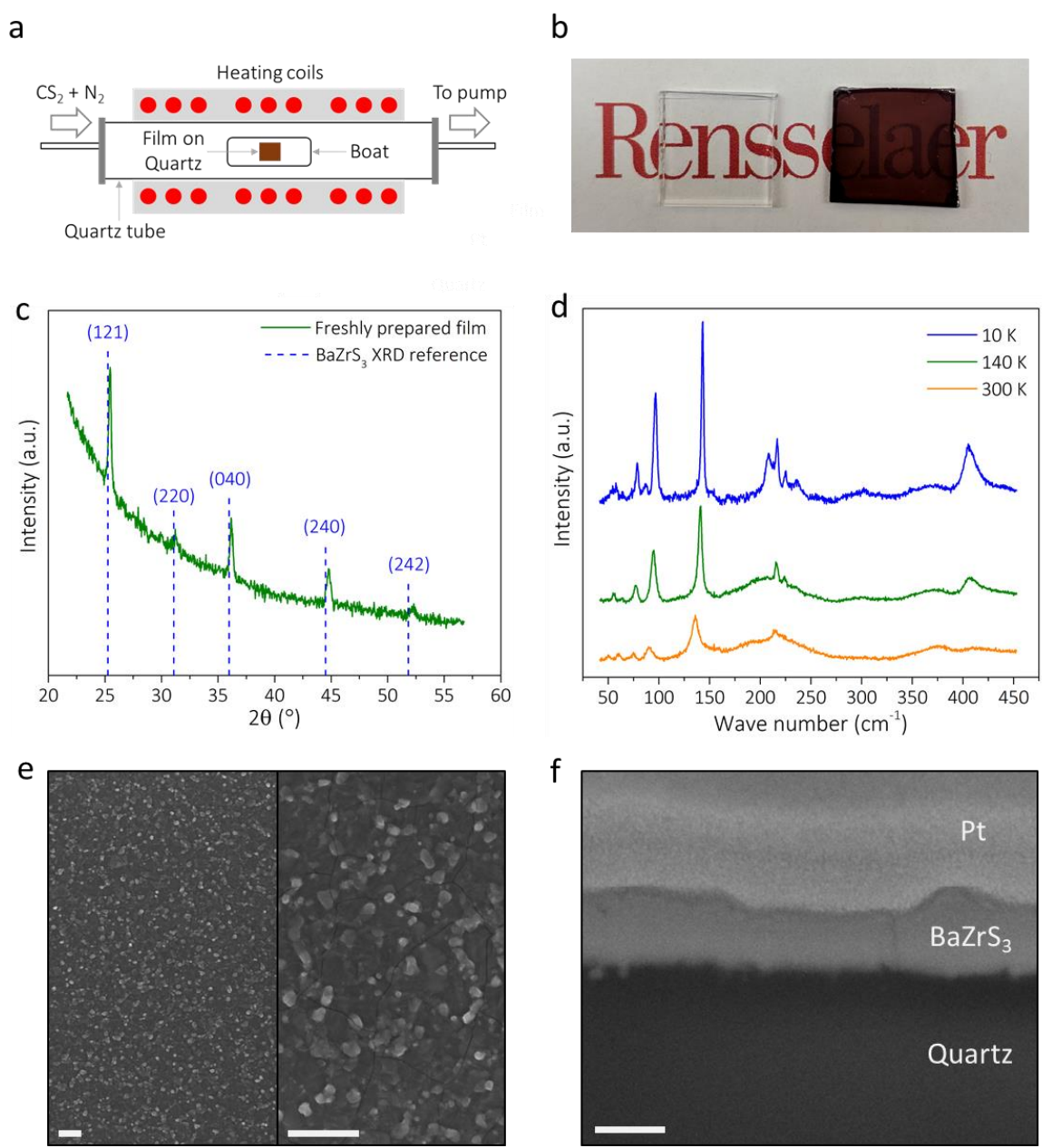

**Figure 1.** Synthesis and characterization of BaZrS$_3$ thin film. **a)** Schematic illustration of the sulfurization process. **b)** Photograph of oxide film (left) and sulfurized film (right) on quartz substrates. **c)** XRD pattern of freshly prepared sulfurized film on quartz substrate. The dashed vertical lines represent lines from the XRD reference for BaZrS$_3$ (ICDD 00-015-0327). **d)** Raman spectra of sulfurized film at 300 K, 140 K, and 10 K. **e)** SEM images of surface of BaZrS$_3$ film. Scale bars are 1 µm. **f)** Cross-sectional SEM image of BaZrS$_3$ thin film on quartz. Scale bar is 200 nm. Pt was deposited to protect the film from FIB damage.

We used Raman spectroscopy to examine the phonon modes of the sulfurized film at 300 K, 140 K and 10 K. The Raman spectra are provided in **Figure 1d**. The peaks sharpened with decrease in temperature due to the expected decrease in anharmonic thermal phonon decay. The observed Raman peaks at 10 K agree well with data for bulk BaZrS$_3$: $A_g^1$ at ~58 cm$^{-1}$, $A_g^2$ and $B_{3g}^1$ at ~78 cm$^{-1}$, $B_{1g}^1$ at ~86 cm$^{-1}$, $A_g^3$ at ~97 cm$^{-1}$, $A_g^4$ at ~143 cm$^{-1}$, $A_g^6$ at ~216 cm$^{-1}$, and $B_{2g}^6$ at ~225 cm$^{-1}$.[21] Through scanning electron microscopy (SEM) imaging (**Figure 1e**), we found that the BaZrS$_3$ thin film was continuous and had a rough surface. A focused ion beam (FIB) was used to create a cross-section and reveal the film's thickness. The cross-sectional SEM image is shown in **Figure 1f**. The mean film thickness was measured to be ~230 nm. Scanning transmission electron microscopy (STEM) provided further evidence that the synthesized film was BaZrS$_3$. A high-resolution HAADF-STEM image overlaid with a crystal model of BaZrS$_3$ is shown in **Figure S2** (Supporting Information). Having confirmed that the deposited material is polycrystalline BaZrS$_3$, we proceeded to investigate the optical characteristics of the BaZrS$_3$ thin film. Its absorption coefficient values and room-temperature photoluminescence (PL) spectrum are provided in **Figure 2a**. A 532 nm laser excitation was used for the PL measurements. The BaZrS$_3$ film exhibited the steepest increase in absorption at a wavelength of ~710 nm, which was also the wavelength of maximum PL intensity. The 710 nm wavelength corresponds to an optical bandgap of ~1.75 eV, which is in good agreement with previously calculated values.[13]

Organic-inorganic halide perovskites (OIHP) lose their optoelectronic potency upon prolonged exposure to moist ambient conditions.[22] Periodically monitoring the PL spectrum has previously been employed to gauge the environmental stability of perovskite materials.[23-24] We followed the same approach to compare the environmental stability of BaZrS$_3$ to the prototypical OIHP – methylammonium lead iodide (CH$_3$NH$_3$PbI$_3$ or MAPbI$_3$). MAPbI$_3$ samples were synthesized by spin-coating a ~25 wt% solution of lead (II) iodide and methylammonium iodide in *N,N*-Dimethylformamide on a 1 cm × 1 cm glass substrate followed by annealing at ~100°C for ~20 minutes. A $BaZrS_3$ thin film and a $MAPbI_3$ thin film were stored together under ambient conditions (~20°C, 40-71% RH) and their PL spectra were monitored. The PL excitation power was kept constant for all measurements. As shown in **Figure 2b**, the PL intensity of $BaZrS_3$ decreased gradually with time but was still substantial after 5 weeks. On the other hand, **Figure 2c** shows that the PL of $MAPbI_3$ vanished after just 2 weeks. **Figure 2d** shows the PL intensity ratios for the $BaZrS_3$ and $MAPbI_3$ samples over time under ambient conditions. $BaZrS_3$ retained ~50% of its initial PL intensity after 5 weeks (the majority of the decay is in the first 1-2 weeks after which the $BaZrS_3$ response tends to level out). By contrast, the PL intensity ratio of $MAPbI_3$ drops precipitously to 0 in 2 weeks. Even more aggressive environmental conditions such as exposure to steam proved to be far less detrimental to $BaZrS_3$ than to $MAPbI_3$. The inset in Figure 2d shows the PL intensity ratios for the two film samples exposed to steam. PL emission from $MAPbI_3$ was extinguished in just 1 minute, but $BaZrS_3$ luminesced with ~82% of its initial PL intensity after 10 minutes of steam exposure. A schematic illustration of the steam exposure experiment and the relevant PL spectra are provided in **Figure S3** (Supporting Information). The PL results demonstrate that the optical response of $BaZrS_3$ is far more stable than that of $MAPbI_3$ in the presence of moisture.

We also established the chemical stability of $BaZrS_3$ by conducting an XRD measurement on the film after extended exposure to atmospheric moisture. **Figure S4a** (Supporting Information) shows that the XRD pattern of the film did not change significantly after 10 weeks in ambient conditions. No new phases were formed and the original $BaZrS_3$ phase persisted. Furthermore, the color of the $BaZrS_3$ film did not change (**Figure S4c**, Supporting Information). A $MAPbI_3$ film under the same conditions underwent rapid degradation. The original $MAPbI_3$ XRD peaks almost vanished after 10 days and prominent peaks of $PbI_2$ appeared (**Figure S4b**, Supporting Information). The formation of $PbI_2$ also explains the color change of the original MAPbI$_3$ film from black to yellow over the 10 day period.

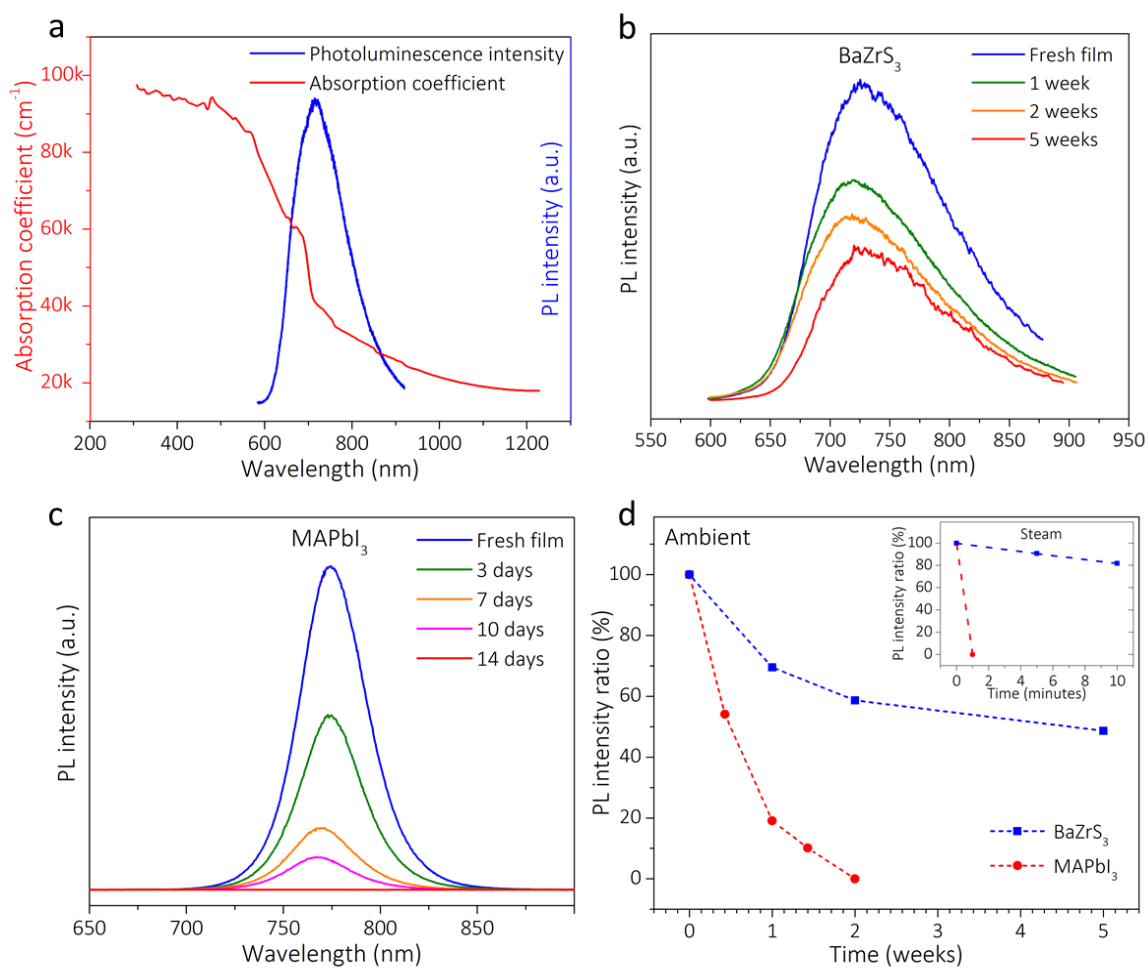

**Figure 2.** Optical characteristics and environmental stability. **a)** Absorption coefficient and photoluminescence (PL) spectrum of BaZrS$_3$ thin film at room temperature. **b)** PL spectra of BaZrS$_3$ thin film kept under ambient conditions for 5 weeks. **c)** PL spectra of MAPbI$_3$ thin film kept under ambient conditions for 2 weeks. **d)** PL intensity ratios of BaZrS$_3$ and MAPbI$_3$ thin films under ambient conditions. The inset shows PL intensity ratios for steam exposure.

We carried out a set of *ab initio* molecular dynamics (AIMD) simulations on BaZrS$_3$ to compare its rate of water-induced degradation with that of MAPbI$_3$. Complementary to the work of Mosconi *et al.* on MAPbI$_3$, the AIMD simulations were based on DFT.[25] Two pristine surfaces of BaZrS$_3$ were relaxed, one with BaS termination and the other with ZrS$_2$ termination. Both surfaces were exposed to a cluster of H$_2$O molecules and the systems were allowed to evolve through AIMD. Throughout the simulations, we tracked the distance between the oxygen atom in $H_2O$ and the surface atoms of $BaZrS_3$. We also tracked the interatomic distances within $BaZrS_3$. As shown in **Figure 3a**, no significant change in these distances was observed for either surface through almost 10 ps of simulation. This behavior is in stark contrast to that of $MAPbI_3$. Data taken from the results of Mosconi *et al.* (**Figure 3b**) shows that the distance between $H_2O$ and Pb decreased as the water rapidly approached the perovskite.[25] $H_2O$ ultimately caused the release of an iodine (I) anion and a neighboring methylammonium (MA) cation. The increased I–Pb and N–Pb distances are reflective of MAI solvation. The unchanging distances in $BaZrS_3$ suggest that its interaction with water is weak. Although the timescale is not fully representative of experimental conditions, our simulations suggest that the rate of water-induced deterioration is much lower for $BaZrS_3$ when compared to $MAPbI_3$.

We also used DFT to examine light-induced degradation. In the case of $MAPbI_3$, the work of Kim *et al.* suggests that photodecomposition involves creation of iodine vacancies in the bulk material.[9] They hypothesize that photoexcitation causes iodine (I) ions in the lattice to absorb holes, allowing them to unbind from the neighboring Pb to form neutral interstitial I atoms. These unbound I atoms have a smaller radius than those bound to Pb, making them highly mobile and likely to diffuse out of the crystal at an accelerated rate. However, our DFT calculations indicate that the interstitial I atoms would invariably bind to nearby Pb atoms when the system was allowed to relax. Considering this, we propose instead that I propagates via migrations between nearest neighbor (NN) I sites. This type of propagation hinges on two processes. First, an I vacancy is created at the $MAPbI_3$ surface due to photoexcitation. This vacancy can then move into the bulk if a neighboring I ion migrates into the vacancy. Through successive migrations, each vacancy takes a random walk, hopping about adjacent I sites and ultimately moving away from the surface and into the bulk material. Over time, such vacancies will "accumulate" in the bulk and alter the material's electronic properties and contribute to the breakdown of $MAPbI_3$. This picture is consistent with Kim *et al.*'s observation of increased $I_2$ outflow under illumination, as I ions that migrate to the surface can detach and join to form $I_2$ molecules. We used the Arrhenius equation to investigate the rates of the aforementioned two processes – i.e., bulk vacancy migration and surface vacancy formation. The Arrhenius relationship can be expressed as:

$$R \propto e^{-E/k_B T} \quad (1)$$

where $R$ is the rate of the given process, $E$ is the energy required to induce the given process, $k_B$ is the Boltzmann constant, and $T$ is the absolute temperature.

For vacancy migration, the energy barrier $E$ was calculated using the climbing image nudged elastic band method (CINEB) by subtracting the energy of the initial equilibrium system from the maximum along the energy profile.[26] Four distinct sulfur vacancy ($V_S$) NN migrations and eight distinct iodine vacancy ($V_I$) NN migrations were examined in $BaZrS_3$ and $MAPbI_3$, respectively. The migration paths are shown in **Figure 3c**. For both materials, the migration barrier energies to second NN sites were ~1 eV higher than those of first NN sites, suggesting that second NN migrations occur at negligible rates at room temperature. The barrier energies for all first NN migrations are plotted in **Figure 3d**. All the barriers for $V_S$ migrations are much higher than those for $V_I$. Quantitatively, the lowest barrier energy for $V_I$ migration was found to be 0.16 eV, which is dwarfed by the lowest $V_S$ migration barrier of 0.59 eV. The Arrhenius equation predicts that at room temperature, a difference in barrier energies of 0.43 eV corresponds to a difference in migration frequency on the order of $10^7$. This is assuming that the vibrational frequencies for modes involving I in $MAPbI_3$ are of the same order of magnitude as those involving S in $BaZrS_3$.[21, 27] Therefore, we expect that the $V_I$ migration rates in $MAPbI_3$ are about seven orders of magnitude higher than those of $V_S$ in $BaZrS_3$. For this reason, surface vacancies in $BaZrS_3$ are not expected to cause instability because the high energy barriers would drastically slow down vacancy migration into the bulk.

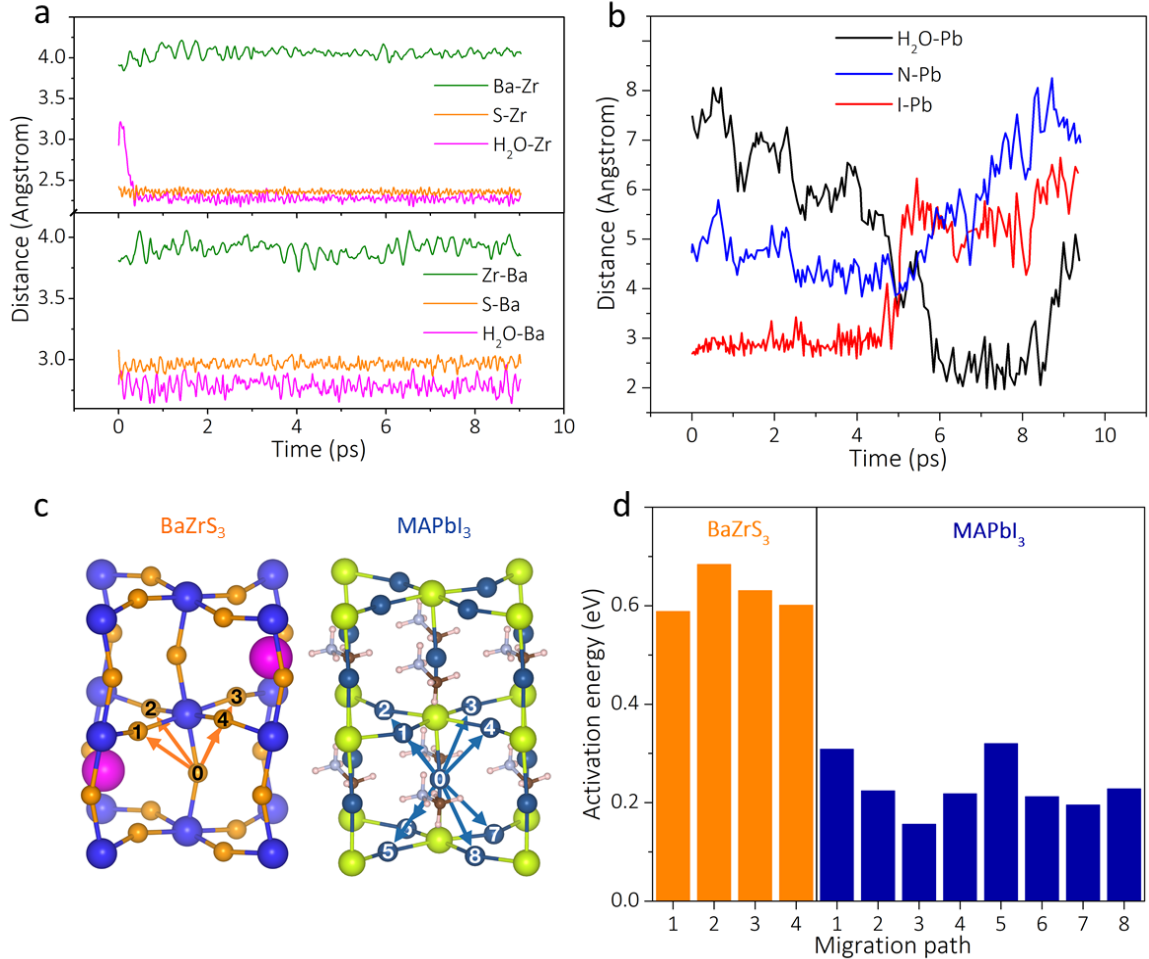

**Figure 3.** Simulations to elucidate BaZrS$_3$'s superior stability. **a)** Interatomic distances throughout the AIMD simulation of BaZrS$_3$ in the presence of water molecules. The lower panel is for BaS surface termination and the upper panel is for ZrS$_2$ surface termination. **b)** Interatomic distances of MAPbI$_3$ in the presence of water molecules. Reproduced with permission.[25] **c)** Anion vacancy migration paths in BaZrS$_3$ and MAPbI$_3$. For BaZrS$_3$, the orange spheres are S atoms, the magenta spheres are Ba atoms, and the blue spheres are Zr atoms. For MAPbI$_3$, the blue spheres are I atoms and the green spheres are Pb atoms. **d)** Barrier energies for sulfur vacancy (orange bars) and iodine vacancy (blue bars) migrations.

We also studied surface vacancy formation in the ground and photoexcited states of MAPbI$_3$ and BaZrS$_3$. Simulation of light-induced processes requires a description of photoexcited systems. For this, the Δ self-consistent field method was used to constrain the occupations of the Kohn-Sham orbitals.[28] To simulate an excited state, the conduction band minimum (CBM) was populated with a single electron, leaving a hole in the valence band maximum (VBM). The energies of the systems of interest were then calculated under these constraints. The formation energy of an iodine vacancy in MAPbI$_3$ is given by:

$$E_f = E_{vac} + \frac{1}{2}E_{I_2} - E_{pris} \qquad (2)$$

where $E_{vac}$ is the energy of MAPbI$_3$ with a surface vacancy, $E_{I_2}$ is that of an I$_2$ molecule, and $E_{pris}$ is that of a pristine MAPbI$_3$ surface. As continuous illumination increases the partial pressure of I$_2$ gas at the MAPbI$_3$ surface, equation 2 assumes that I atoms that leave MAPbI$_3$ enter an iodine-rich environment. As indicated in **Table S1** (Supporting Information), the surface vacancy formation energies are much smaller in the excited system than in the ground state system. The smaller energies would enable rapid V$_I$ creation when MAPbI$_3$ is subjected to photoexcitation. Rapid V$_I$ migration coupled with accelerated surface V$_I$ creation explains the breakdown of MAPbI$_3$ under photoexcitation. We found that photoexcitation also lowers surface sulfur vacancy formation energies for BaZrS$_3$ (Table S1, Supporting Information). However, BaZrS$_3$ is relatively immune to photodecomposition due to its much lower V$_S$ migration rates.

To test the viability of BaZrS$_3$ for optoelectronics, we fabricated and characterized photodetectors of the lateral photoconductor type. Square-shaped contact pads of gold were deposited on top of BaZrS$_3$ thin films to make the devices. The pads were ~60 nm thick and ~425 µm long with a spacing of ~83 µm. Similar devices were fabricated with MAPbI$_3$ thin films for comparison. Devices of both materials were stored together under ambient conditions (~20°C, 40-71% RH) and current-voltage (I-V) characteristics of these devices were measured periodically. Measurements were made in the dark and under illumination with a 405 nm laser. The illumination power density was ~55 mW cm$^{-2}$ for all measurements. A schematic illustration of a device and its measurement is provided in **Figure 4a**. The I-V characteristics of a BaZrS$_3$ photodetector are shown in **Figure S5** (Supporting Information). The linear I-V relationship shows that the contact between BaZrS$_3$ and gold was ohmic. The dark current of BaZrS$_3$ photodetectors was substantial, which may be because of defects. The photocurrent was calculated by subtracting dark current from illuminated current. Responsivity (A W$^{-1}$) was calculated by dividing the photocurrent density (A cm$^{-2}$) by the illumination power density (W cm$^{-2}$).

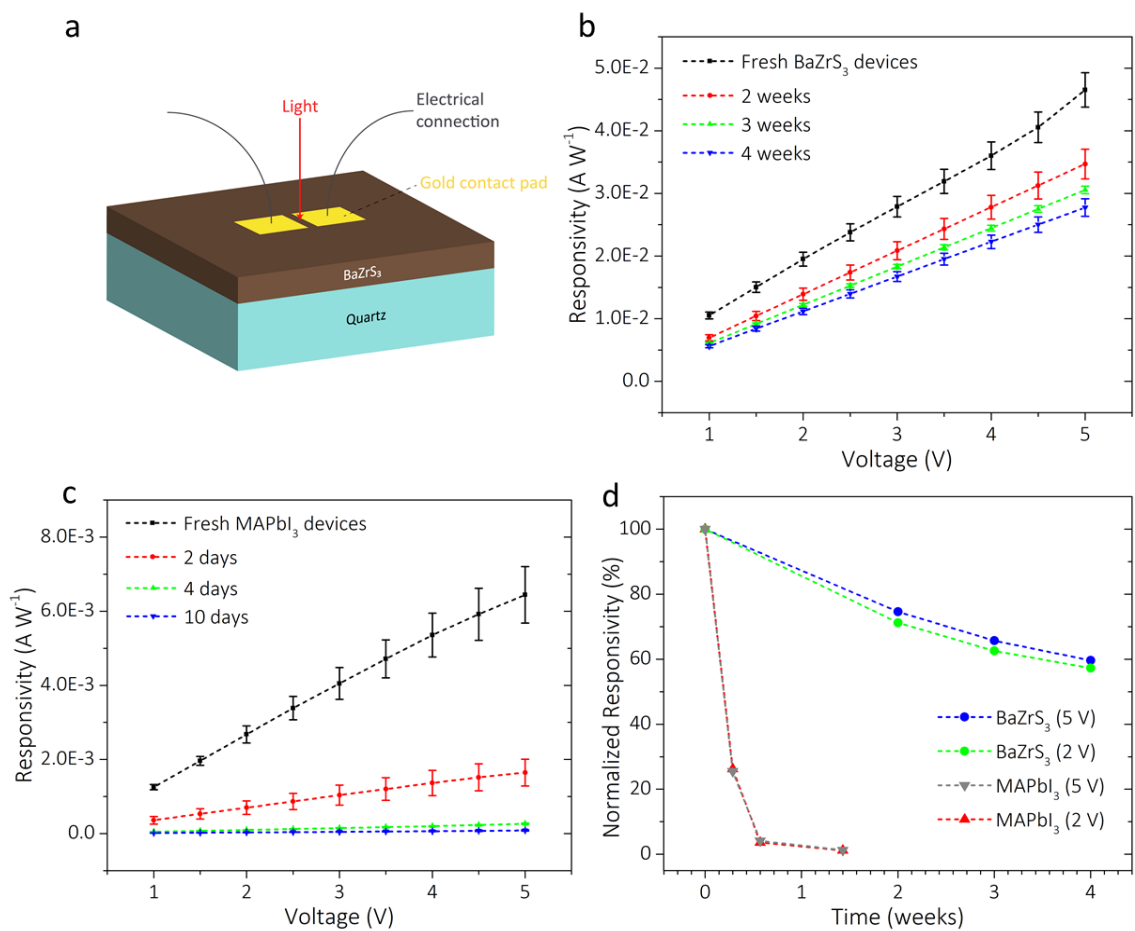

**Figure 4.** Photodetector characterization and performance. **a)** Schematic illustration of BaZrS$_3$ photodetector. **b)** Responsivity of BaZrS$_3$ photodetectors kept under ambient conditions for 4 weeks. **c)** Responsivity of MAPbI$_3$ photodetectors kept under ambient conditions for 10 days. **d)** Normalized responsivity of BaZrS$_3$ and MAPbI$_3$ photodetectors kept under ambient conditions. All responsivity measurements were performed with 405 nm laser illumination at a power density of ~55 mW cm$^{-2}$.

Responsivity values of the BaZrS$_3$ photodetectors are shown in **Figure 4b**. Fresh BaZrS$_3$ devices exhibited a mean responsivity of ~46.5 mA W$^{-1}$ at 5 V. This value is comparable to the reported responsivities for lateral polycrystalline OIHP photodetectors illuminated with similar wavelengths. Wang *et al*. reported ~17.5 mA W$^{-1}$ with 400 nm at 10 V and Hu *et al.* reported ~110 mA W$^{-1}$ with 470 nm at 3 V.[29,30] Responsivity values of the BaZrS$_3$ devices after 2, 3, and 4 weeks in the ambient are also shown in Figure 4b. For comparison, responsivity values of the MAPbI$_3$ devices are shown in **Figure 4c**. A mean responsivity of ~6.4 mA W$^{-1}$ was observed for fresh MAPbI$_3$ devices at 5 V, but the responsivity slumped by two orders of magnitude in just 10 days. MAPbI$_3$ also changed color from black to yellow in that time, indicating decomposition to PbI$_2$. Photographs of a MAPbI$_3$ device before and after degradation are shown in **Figure S6** (Supporting Information). BaZrS$_3$, as observed previously, did not change color in 4 weeks. Normalized responsivity values of the BaZrS$_3$ and MAPbI$_3$ devices are shown in **Figure 4d**. In just 4 days, the mean responsivity of the MAPbI$_3$ devices at 5 V decreased to ~5% of the initial value. In 10 days, the photo-responsivity of MAPbI$_3$ is nearly completely extinguished. On the other hand, the BaZrS$_3$ devices exhibited a much more stable photo-response by retaining ~60% of the initial responsivity at 5 V after 4 weeks in the ambient. The majority of the aging occurs in the first 1-2 weeks after which the BaZrS$_3$ film's photo-responsivity tends to level off. Responsivities at other voltages followed a similar trend.

In addition to optoelectronics, we also show the potential of this material for energy conversion in thermoelectric devices by using first principles calculations (see Materials and Methods). The extremely low thermal conductivity of BaZrS$_3$ contributes to an enhanced thermoelectric figure of merit (*ZT*) over a wide range of temperatures (**Figure S7**, Supporting Information). At higher temperatures (500 - 700 K), *ZT* remains constant (maximum ≈ 1) over a wide range of carrier concentrations. Here, *ZT* is dominated by the Seebeck coefficient until the peak carrier concentration of 10$^{18}$ cm$^{-3}$ is attained. Further increasing the carrier concentration causes a higher entropic contribution to the thermoelectric energy conversion process that leads to a decrease in *ZT*. At lower temperatures (< 500 K), *ZT* decreases with carrier concentration because combined effect of the Seebeck coefficient and electrical conductivity leads to higher carrier scattering.

To summarize, we report a chalcogenide perovskite thin film and photodetector. This perovskite – BaZrS$_3$ – was synthesized by sulfurizing a BaZrO$_3$ thin film. The BaZrS$_3$ thin film was found to be polycrystalline with a bandgap of ~1.75 eV. The $BaZrS_3$ film substantially outperformed $MAPbI_3$ in terms of stability under moisture-rich conditions. Our simulations indicated that $BaZrS_3$ interacts very weakly with water when compared to that of $MAPbI_3$. Our calculations also showed the rate of anion vacancy migration in $BaZrS_3$ to be seven orders of magnitude slower than that in $MAPbI_3$, making $BaZrS_3$ far less prone to photodecomposition. The advantage of environmental stability was seen clearly in photodetector performance. $BaZrS_3$ photodetectors lost ~40% of their initial responsivity after 4 weeks in the ambient, whereas similar $MAPbI_3$ photodetectors degraded by ~95% in only 4 days. Our results provide experimental evidence and theoretical explanations for the environmental stability of $BaZrS_3$. Lack of toxic lead (Pb) and intrinsic stability under photoexcitation and when exposed to the environment makes this chalcogenide perovskite a viable candidate for optoelectronics. The material also shows promise as a high figure of merit material for thermoelectric energy conversion. Future efforts with $BaZrS_3$ should focus on lowering the synthesis temperature and reducing the dark current in devices.

**Materials and Methods**

*Synthesis of $BaZrO_3$ thin film:* 1.92 g of barium acetate (99%, Alfa Aesar), 3.66 g of zirconium(IV) acetylacetonate (97%, Sigma-Aldrich) and 0.90 g of polyvinyl butyral (Sigma-Aldrich) were stirred and dissolved in 25 mL of propionic acid (99.5%, Sigma-Aldrich) at 60°C. The resulting clear and transparent solution was spin-coated on a clean quartz substrate (1 cm × 1 cm × 2 mm) at 2000 rpm for 1 minute followed by 5000 rpm for 5 minutes. The spin-coated film was annealed in air in a Thermolyne FB1315M muffle furnace at 700°C for 15 minutes followed by 40 minutes at 870°C.

*Synthesis of $BaZrS_3$ thin film:* The $BaZrO_3$ thin film on quartz was placed in a quartz boat in the middle zone of an MTI OTF-1200X three-zone tube furnace (quartz tube with 3″ diameter). The tube was evacuated down to a base pressure of ~30 mTorr and then purged with UHP nitrogen (N$_2$) while maintaining a pressure of ~150 mTorr. All three zones were then ramped up to 1050°C in 1 hour. When the temperature reached 600°C, carbon disulfide (CS$_2$) was introduced into the tube through a bubbler filled with liquid CS$_2$ (99.9%, Sigma-Aldrich). UHP N$_2$ was used as the carrier gas and the bubbler was kept at ~20°C. A mass flow controller at the outlet of the bubbler was used to keep the flow rate of the CS$_2$-N$_2$ mixture at ~25 sccm while a pressure of ~2 Torr was maintained inside the tube. The furnace was held at 1050°C for 4 hours. Then the heating was stopped, and the furnace was allowed to cool down naturally without opening the lid. When the furnace had cooled down to 600°C, the CS$_2$ supply was stopped and the tube was purged with UHP N$_2$ till the furnace cooled down completely. The tube was then brought up to atmospheric pressure and the sulfurized film was extracted.

*Synthesis of MAPbI$_3$ thin film:* 461 mg of lead (II) iodide (99.999%, Sigma-Aldrich) and 159 mg of methylammonium iodide (99%, Sigma-Aldrich) were dissolved in 2 mL of *N,N*-Dimethylformamide (99.9%, EMD Millipore) to obtain a clear solution. This solution was spin-coated on a clean glass substrate (1 cm × 1 cm × 1 mm) at 2000 rpm for 30 seconds. Finally, the spin-coated film was annealed on a hot plate at 100°C for 20 minutes. All the steps in this synthesis were carried out in an argon-filled glovebox.

*Materials characterization*: X-ray diffraction (XRD) measurements were conducted on a PANalytical X'Pert Pro diffractometer using CuKα ($\lambda$ =1.5405 Å) radiation. The X-ray generator was set to 45 kV and 40 mA. Raman spectra were acquired with a Horiba Jobin-Yvon LabRAM HR evolution Raman spectrometer in back scattering geometry with 633 nm laser excitation and a Peltier-cooled CCD detector. A Carl Zeiss 1540EsB Crossbeam system was used for scanning electron microscopy (5 kV) and focused ion beam work (Ga ion, 30 kV). Photoluminescence spectra were acquired by using 532 nm laser excitation and an Andor spectrograph with a Peltier-cooled CCD detector. The laser power was measured by using a power meter. Transmission electron microscopy was carried out on a FEI Titan cubed STEM equipped with a monochromator and probe corrector. The HAADF detector was used and imaging was performed in STEM mode at 300 kV with a 0.5 nA beam current.

*Device fabrication and characterization:* 60 nm thick gold contacts were deposited on the $BaZrS_3$ and $MAPbI_3$ thin film samples by e-beam evaporation at a deposition rate of ~1 Å/s. A copper shadow mask was used to create the pattern. The current-voltage characteristics were measured by using a Keithley 4200-SCS semiconductor characterization system in a two-probe configuration. A 405 nm laser was used for photoexcitation. The laser power was measured by using a power meter.

*Computational methods:* DFT calculations were carried out using the projector augmented wave (PAW) method of density functional theory (DFT) implemented in the Vienna ab initio simulation package (VASP).[31-35] The Perdew-Burke-Ernserhof (PBE) generalized gradient approximation (GGA) for the exchange-correlation functional was employed,[36] with a basis set including plane waves with energies up to 400 eV. The Brillouin zones of pristine $MAPbI_3$ and $BaZrS_3$ were respectively sampled with 4×4×3 and 3×3×2 Γ-centered Monkhorst-Pack grids.[37] Relaxation iterations continued until the Hellmann-Feynman forces on all atoms settled below 10 meV/Å, while electron field iterations persisted until changes in both the total energy and Kohn-Sham eigenvalues fell below $10^{-5}$ eV. For relaxations of surface structures, 12 Å of vacuum was inserted in the z-direction (out-of-plane direction) to ensure that interactions with the periodic images were negligible. AIMD and CINEB simulations obeyed the same convergence criteria. AIMD simulations were run in an NVT ensemble at a temperature of 300 K with a 1 fs time step. To simulate the perovskite surfaces, 2×2 slabs were cut from the $BaZrS_3$ crystal, which exposed the BaS- and $ZrS_2$-terminated (001) surfaces after relaxation. The vacuum regions above and below the perovskite slabs were populated with water molecules, whose density was kept consistent with the experimental density of liquid water. The CINEB calculations included eight image structures along each ionic path. Care was taken to remove metastable states from any ionic path before attempting to relax it. That is, any path that contained a metastable state was split into two paths, each bounded on one end by that metastable state. Thermoelectric properties were calculated using the linearized Boltzmann transport equations in the relaxation time approximation using a Fourier expansion of the electronic energies as obtained from VASP for the optimized structures. The thermoelectric figure of merit $ZT = \frac{S^2 \sigma T}{\kappa_e + \kappa_L}$, where $S$ is Seebeck coefficient, $\sigma$ the electrical conductivity, $T$ the temperature, while $\kappa_e$ and $\kappa_L$ are the electronic and lattice thermal conductivities, respectively. A denser *k*-mesh of 120,000 points was employed to ensure higher accuracy of the calculated transport properties.

**Acknowledgements**

T.G. is grateful to Kent Way and Bryant Colwill for building and maintaining the sulfurization setup, to David Frey for helping with the FIB, and to Vidhya Chakrapani for helping with absorption spectroscopy. N.K. and V.M. acknowledge funding support from the USA National Science Foundation (Award 1608171). S.-F.S. acknowledges support from AFSOR through Grant FA9550-18-1-0312. G.B. thanks support from the P.C. Rossin Assistant Professorship at Lehigh.

# Supporting Information

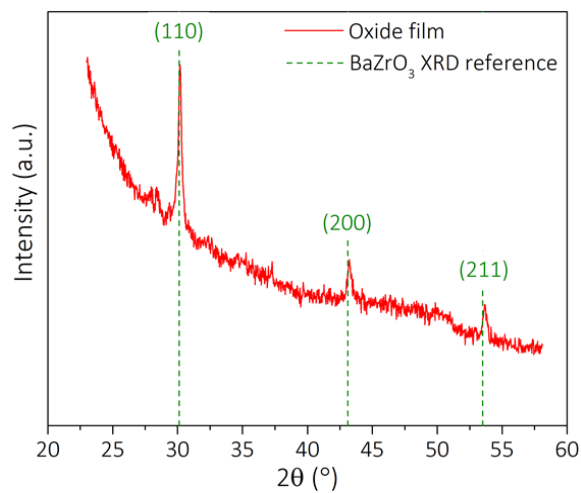

**Figure S1.** XRD pattern of BaZrO$_3$ oxide film on quartz substrate. The dashed vertical lines represent lines from the XRD reference for BaZrO$_3$ (ICDD 00-006-0399).

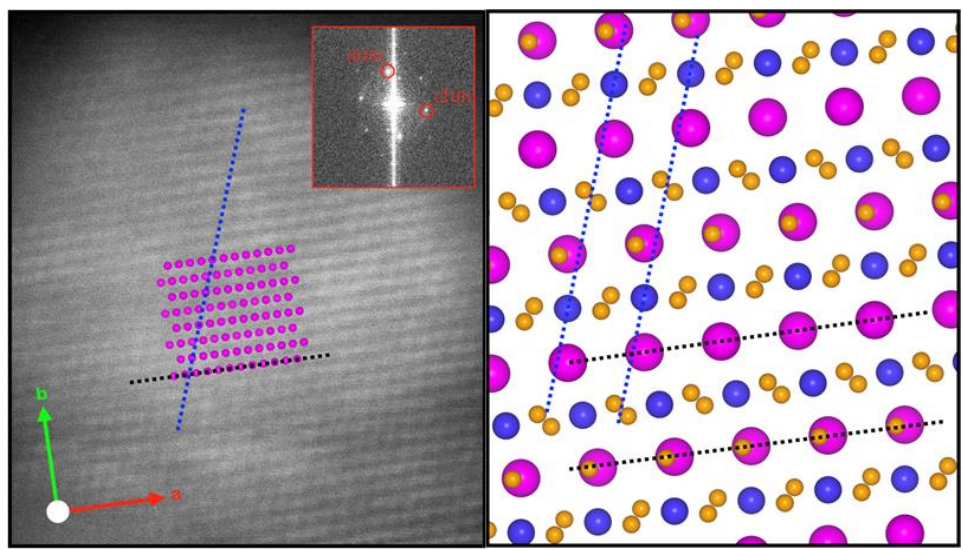

**Figure S2.** HAADF-STEM image of BaZrS$_3$ thin film (left). The overlay is from the atomic model on the right and shows the position of Ba atoms. The inset shows the fast Fourier transform of the STEM image. On the right, the simulated BaZrS$_3$ structure is shown, where magenta spheres are Ba atoms, orange spheres are S atoms, and blue spheres are Zr atoms.

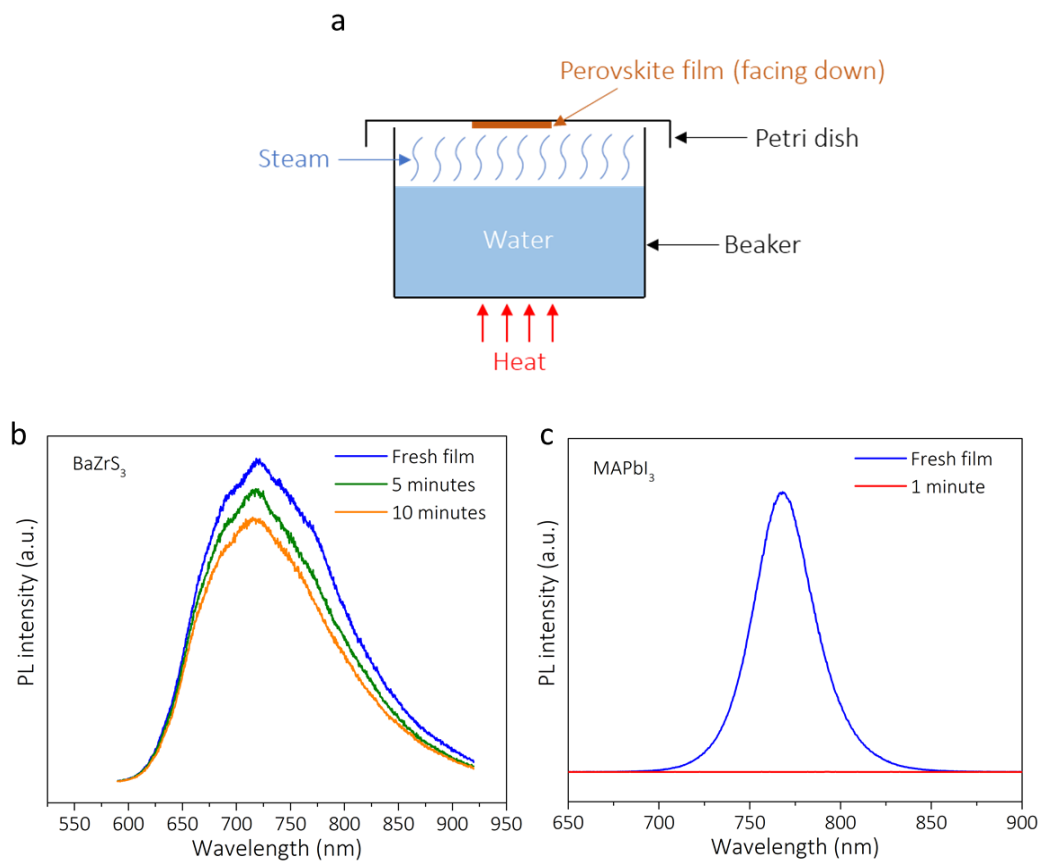

**Figure S3.** (**a**) Schematic illustration of steam exposure. (**b**) PL spectra of BaZrS$_3$ thin film exposed to steam. (**c**) PL spectra of MAPbI$_3$ thin film exposed to steam.

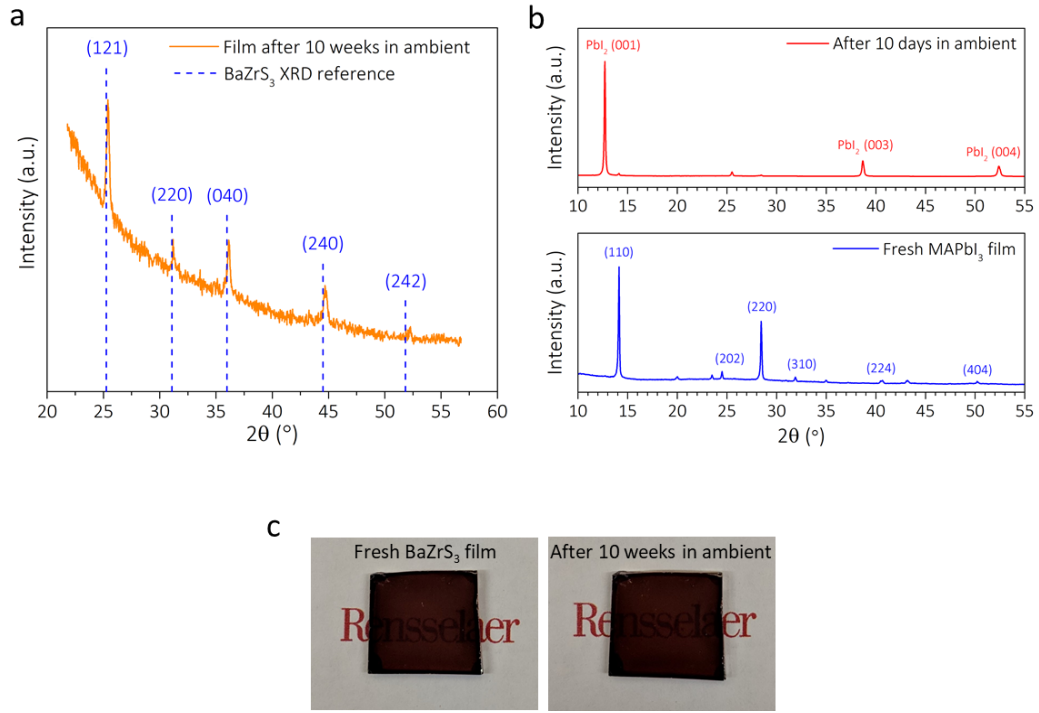

**Figure S4. (a)** XRD pattern of BaZrS$_3$ thin film after 10 weeks under ambient conditions. **(b)** XRD patterns of fresh (lower panel) and degraded (upper panel) MAPbI$_3$ thin film. **(c)** Photographs of a BaZrS$_3$ thin film on quartz substrate.

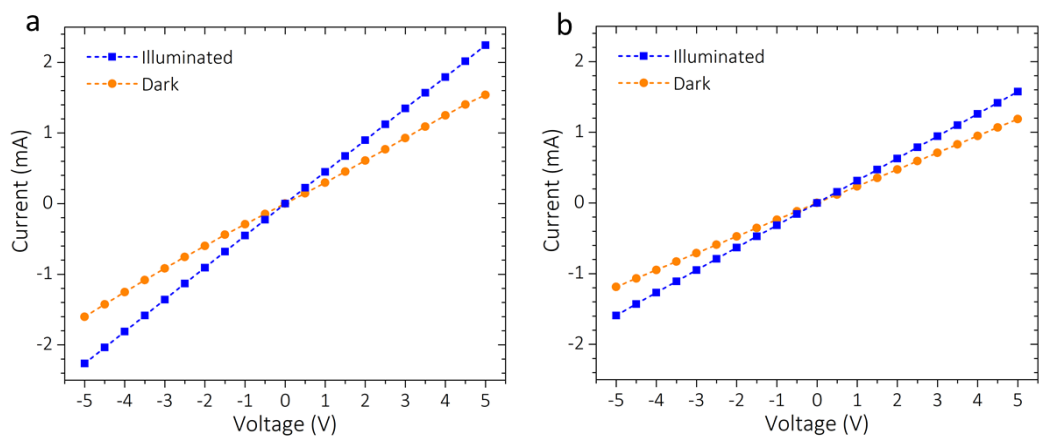

**Figure S5. (a)** Current-voltage characteristics of a fresh BaZrS$_3$ photodetector. **(b)** Current-voltage characteristics of the same BaZrS$_3$ photodetector after 4 weeks in ambient conditions.

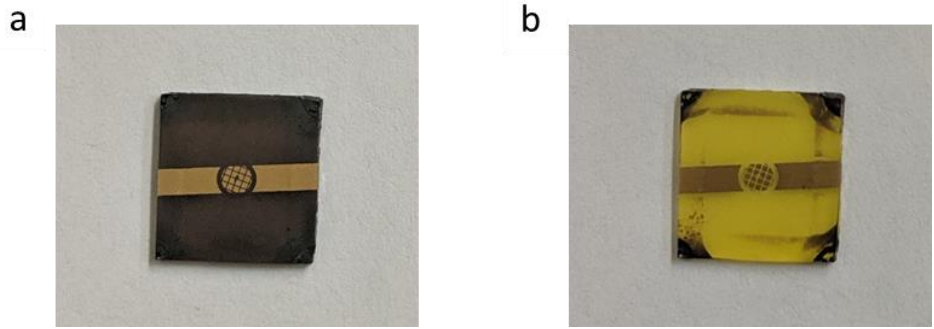

**Figure S6. (a)** Photograph of a fresh MAPbI$_3$ photodetector. The deposited gold contact pads can be seen on top of the MAPbI$_3$. **(b)** Photograph of the same MAPbI$_3$ photodetector after 10 days in the ambient.

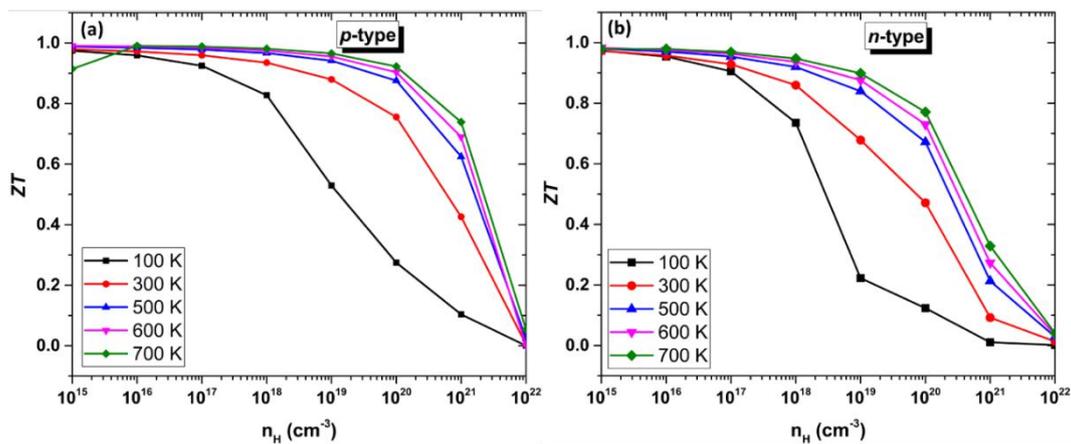

**Figure S7.** Calculated thermoelectric figure of merit (*ZT*) against carrier concentration *n* for **(a)** *p*-type doping and **(b)** *n*-type doping across a temperature range for BaZrS$_3$.

**Table S1.** Surface iodine and sulfur vacancy formation energies for MAPbI$_3$ (for PbI$_2$ and MAI surface terminations) and BaZrS$_3$ (for ZrS$_2$ and BaS surface terminations).

|  | PbI$_2$ | MAI | ZrS$_2$ | BaS |
| --- | --- | --- | --- | --- |
| Ground (eV) | 1.5425 | 2.2343 | 0.2732 | 0.4783 |
| Excited (eV) | 0.5624 | 0.3872 | 0.1005 | -0.5586 |